\documentclass[preprint]{elsarticle}

\usepackage[colorlinks=true, citecolor=blue]{hyperref}
\usepackage{amsmath}
\usepackage{aas_macros}
\usepackage{caption}
\usepackage{tikz}
\usepackage{nicefrac}

\newcommand{\word}[3]{
	\{#1\}_{#2}^{(#3)}%
}

\usetikzlibrary{shapes.geometric}
\usetikzlibrary{decorations}
\usetikzlibrary{decorations.pathreplacing}
\usepgflibrary{arrows.meta}

\journal{Astronomy and Computing}

\newcommand\fips{\textsc{Fips}}
\newcommand\opengl{{OpenGL}}

\begin{document}

\begin{frontmatter}

\title{Fips: an OpenGL based FITS viewer}

\author{Matwey Kornilov}
\cortext[mycorrespondingauthor]{Corresponding author}
\ead{matwey@sai.msu.ru}
\author{Konstantin Malanchev}
\address{Sternberg Astronomical Institute, Lomonosov Moscow State University\\ Universitetsky pr. 13, Moscow 119234, Russia}
\address{National Research University Higher School of Economics\\
\nicefrac{21}{4} Staraya Basmannaya Ulitsa, Moscow 105066, Russia}

\begin{abstract}
{
FITS~(Flexible Image Transport System) is a common format for astronomical data storage. It was first standardised in the early 1980s~\cite{Wells1981}.
Even though astronomical data is now processed mostly using software, visual data inspection by a human  is still important during equipment or software commissioning and while observing.
We present \fips{}\footnote{\url{https://fips.space}}, a cross-platform FITS file viewer open source software~\cite{Fips}.
To the best of our knowledge, it is for the first time that the image rendering algorithms are implemented mostly on GPU~(graphics processing unit).
We show that it is possible to implement a fully-capable FITS viewer using \opengl{}~\cite{opengl}
interface. We also emphasise the advantages of using GPUs for efficient image handling.
}
\end{abstract}


\end{frontmatter}

\section{Introduction}

FITS~(Flexible Image Transport System), a famous image format, was first introduced a few decades ago, in the early 1980s~\cite{Wells1981}.
Since that time, it has become the most popular format to store astronomical optical observations.
Now, in 2019, FITS looks like a legacy format rather than a modern technology item~\cite{scroggins_boscoe2018}.
For instance,
it uses the big-endian storage format, while most current server, desktop and mobile
 processors use little-endian storage format~\cite{intel_x86_2015,windows_arm_2016,apple_ios_2013}; moreover, FITS relies internally  on a 2880-byte-alignment structure~\cite{Wells1981}.
Whereas it was a natural choice for tape-based storage media, modern file systems on hard drives and solid state drives operate with pages of $2^N$ size.
For some upcoming projects, it is the JPEG~2000~\cite{skodras_etal2001} data format that is being discussed now by the astronomical community~\cite{Kitaeff2015}.

However, there are two main things that make us believe that FITS will continue be used for decades to come.
First, bulk of astronomical data in the world is stored in FITS format.
Second, astronomical data acquisition software or data processing software are mostly FITS-centric.
As long as visual human inspection of raw or processed astronomical image data  is still an important part in algorithm and software troubleshooting procedures, as well as in hardware alignment and commissioning, it would be helpful to have yet another FITS image viewer software.

There is, of course, a lot of available software doing this work well, among them SAO~DS9~\cite{ds9_2000}, Ginga~\cite{jeschke_etal2013,jeschke_2013}, etc.
All these solve the same problems while rendering data from a file to the user screen.
An application has to parse and read the data from the file.
Then, it has to make geometric transformations for scale, pan, rotation, etc.
After that, colour maps are applied in order to narrow down the FITS range of values to the 8 bit colour depth.

In this paper we concentrate on how these tasks may be offloaded to the GPU~(graphics processing unit).
Modern GPUs have many hard-wired features accelerating typical 2D and 3D-rendering tasks.
These advantages are employed by a wide variety of desktop applications such as web-browsers~\cite{Rouget2010,Wiltzius2014} and even graphical terminal emulators~\cite{Wilm2017,Nachman2018}.
We propose a FITS viewer software implementation based on the GPU acceleration.
After a raw FITS file data loaded into the GPU memory, the geometric and colour transformations are handled by GPU.

It is widely believed by the developer community that software generally should use a minimal amount of the computational resources while keeping its source code simple, clear, and easily maintainable.\footnote{Many recognised experts in computer science often note this. For instance, Cormen et al. (\cite{Cormen2009},~\S~I.2) say: ``If computers were infinitely fast, any correct method for solving a problem would do. You would probably want your implementation to be within the bounds of good software engineering practice...''; or Knuth (\cite{Knuth1968},~\S~1.1) says: ``In practice we not only want algorithms, we want good algorithms in some loosely-defined aesthetic sense. One criterion of goodness is the length of time taken to perform the algorithm... Other criteria are the adaptability of the algorithm to computers, its simplicity and elegance, etc.''.}
The GPU-based end-to-end rendering of the raw FITS file data fulfils both of this conditions.
The specific dedicated design of modern GPUs provides an opportunity to render graphics in energy and computationally efficient way.
GPU programming interfaces such as OpenGL, Direct~X, or Vulkan allow a programmer to describe the high-level geometric and colour characteristics of the scene.
In our case, only a few lines of high-level source code are required to program GPU.

\begin{figure}[t!]
 \centering
 \includegraphics[width=0.8\linewidth]{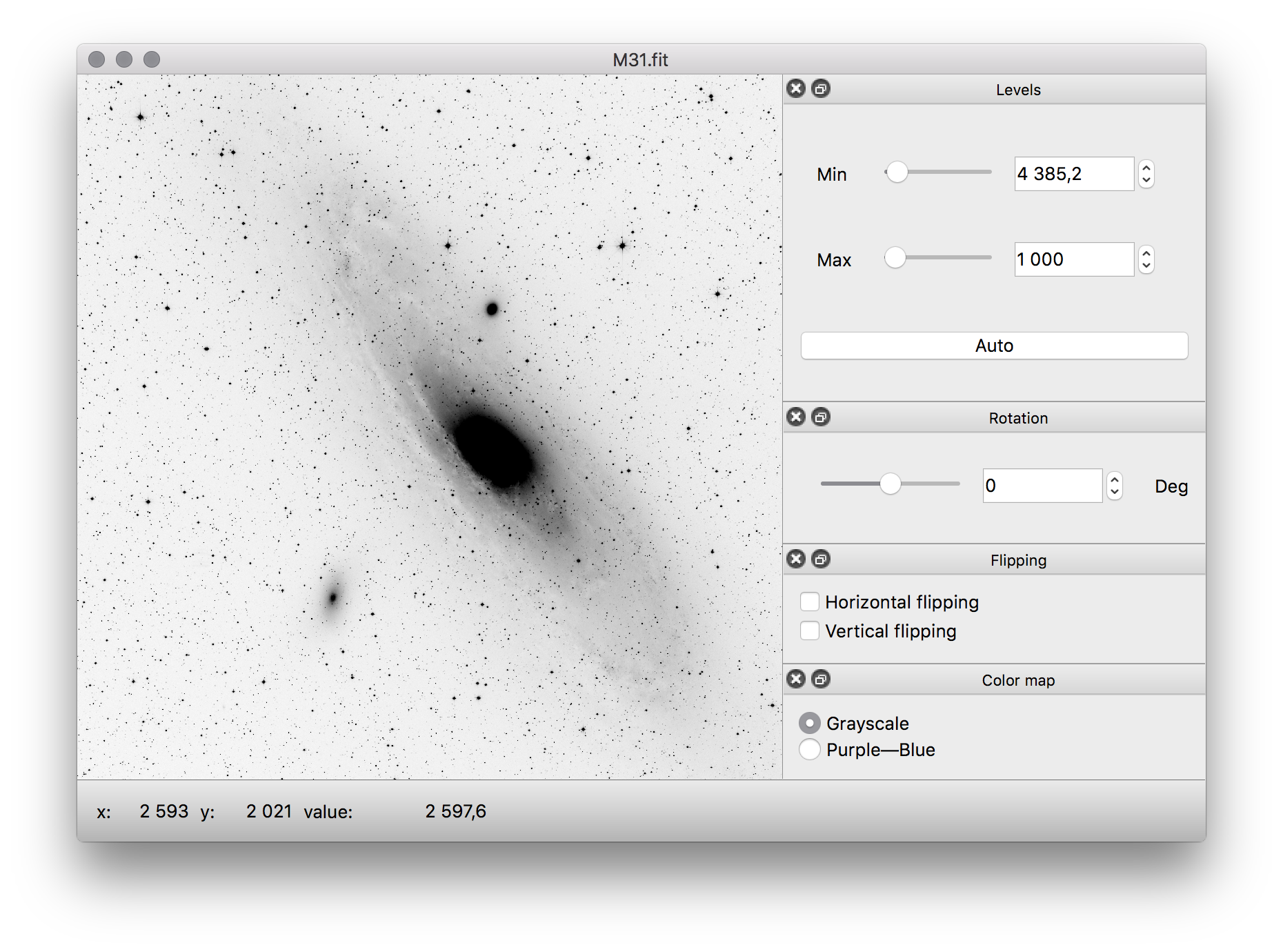}
 \caption{\label{fig:screenshot}\fips{} interface on the macOS operating system. The user interface looks the same both on Linux and Windows.
A M31 galaxy image obtained by the MASTER robotic telescopes network~\cite{Master2010} is shown here.
}
\end{figure}

In this paper we prove that the end-to-end OpenGL rendering of FITS file data can be practically implemented as software application.
This proof of concept application is called \fips{}.
It is a cross-platform open-source graphical user interface software.
The application main window is shown in Fig.~\ref{fig:screenshot}.





The outline of the paper is as follows.
Section~\ref{s.opengl} explains briefly how modern graphics processing units operate and how one could control them using the \opengl{} interface.
Section~\ref{s.fits} describes the essence of the FITS format and how it can interoperate with the \opengl{}.
Then, in Section~\ref{s.storage}, we demonstrate how the FITS data may be transformed into a coloured image displayed by \fips{}.
The software evaluation and testing are described in Section~\ref{s.eval}.
In the \hyperref[s.discussion]{Discussion} section, we highlight further possible applications of the techniques described here.
Finally, we summarise the advantages of the GPU use in the \hyperref[s.conclusion]{Conclusion}.
\ref{s.coordinate} describes how the FITS and \opengl{} coordinate systems are connected to each other.



\section{Brief Introduction to \opengl{}}
\label{s.opengl}
General-purpose graphics processing units are widely used in modern astronomy for solving visualisation problems~\cite{perkins_etal2014} and computational linear algebra problems~\cite{Fromang2006,Liska2018},
with the latter seemingly more widespread. Thus, it makes sense to briefly introduce the graphics pipeline here.
A more comprehensive \opengl{} description can be found in its specification~\cite{opengl}.
In what follows, only details that are important for current work will be highlighted.

\opengl{} is essentially a programming interface, i.e. a library with a standardised set of functions available on different platforms or operating systems.
It is assumed (although not required) that the library functions should be implemented using the graphics processing unit hardware, which is capable to perform a limited set of the most frequently required graphics rendering operations.

Generally, the programming work that uses the GPU with \opengl{} does not resemble conventional programming; instead, one may substitute routines of some kinds into the predefined data processing pipeline and operate with the predefined types of data objects. 

In an \opengl{} application, different kinds of objects may exist.
A point in 3D space that belongs to a geometric object is called a {\it vertex}.
In typical 3D applications, geometric objects normally consist of hundreds and thousands of vertices; however, for our purposes, it is sufficient to have four vertices forming a rectangular plane in our application.
The plane is used to draw picture on its surface.
Arrays of vertex attributes~(coordinates) are transferred from CPU to GPU.

It is usually convenient to define vertex coordinates in an object-centric coordinate system. 
However we need to place each object at its desired place in order to form the world scene.
This is the reason for {\it vertex shaders}.
These are GPU-side routines (functions) performing an arbitrary geometric transformation of vertex coordinates.
The shaders are designed to process only one vertex at a time.
Since there is no concurrence that allows engaging in parallel as many shader units during the execution as are available at one particular hardware.
They are usually used to transform vertex coordinates from the local object coordinate system to the camera coordinate system.
The transformation is often defined as a matrix-vector operation; these transformations may be of various kinds: we can translate, rotate, scale or distort an object.
In fact, what the user can see at the display is a world projection onto the xy coordinate plane.

Graphical memory fragments that contain raster pictures are called {\it textures}.
They are similar to traditional uncompressed bitmap images and FITS images.
The texture content also may be transferred between CPU and GPU in both directions.
The possible ways to arrange the pixel colours in texture memory are given in the \opengl{} specification~\cite{opengl}.
For instance, one may store a colour picture as an array of usual RGB triplets and a monochromatic image as an array of single channel pixels.
The process of transforming FITS images into \opengl{}  for all possible formats will be discussed in  Section~\ref{s.storage}.

Texture data are accessed via {\it samplers}.
A sampler can be thought of as the C {\tt operator []} indexed by a real value coordinate rather than by an integer index.
Among other things, samplers perform the interpolation before returning a value.
There may be two kinds of interpolation: the nearest neighbour interpolation and the linear interpolation.
Although we use the nearest neighbour interpolation, the returned result is always normalised to $[0;1]$ range independently of the data storage format.\footnote{\opengl{}~3.0 and later versions support the access to an normalised raw value. As we see later in Section~\ref{s.storage}, this does not give us a great advantage}
For instance, the standard 8-bit colour channel value $255$ stored in a texture will be returned as $1.0$ by a sampler.

Vertices form the edges and faces of geometric objects, and textures can be drawn on these faces.
At the same time, the orientation between the surface and the camera is also taken into account; for instance, the hidden parts of objects are not rendered at all.

{\it Fragment shaders} are GPU routines that are used to colourise sides of geometric objects.
Formally, this can be considered as the mapping between fragment coordinates and their required colour, where
each fragment is a tiny piece of the scene seen by the user.
In our simple case, the fragment shader prescribes that the hardware would draw the texture containing the image onto a singleton rectangular plane without distortion.

After everything is loaded to the graphics memory and preconfigured, an application would trigger the drawing, and the user would be able to see 3D scene snapshot at the screen.
In case the objects are to be rearranged, altered vertex shaders parameters are loaded and the drawing is performed again.
In modern 3D professional software and games, it is usually required to perform multiple renderings to achieve realistic light distribution over the single frame.
For this purpose, depth textures and stencils are used.

\section{FITS Rendering Implementation}\label{s.fits}
Our goal could be achieved by performing two tasks.
First, we need to load a FITS image to the graphics memory.
Second, we should  render the image using the required orientation, magnification, colour-mapping, etc.

If any transformation between the FITS image memory representation and GPU texture representation is needed, it would use additional CPU-based calculations.
This is why we focus on cases where byte-representations are similar at CPU (FITS) and GPU (texture) sides.
For instance, a 16-bit FITS image is an array of subsequent 16-bit integers.
A monochrome 16-bit texture is also an array of subsequent 16-bit integers.
Since it is monochrome (i.e. single-channel), a single integer represents a whole single pixel.
This means that memory representations of  the 16-bit FITS image and the 16-bit single channel texture should be identical except for the byte order (see below).
Unfortunately, this is not always the case.
For instance, if we have a 64-bit integer FITS image, we cannot just specify that the texture should have a single 64-bit channel, because there are no 64-bit integer textures in \opengl{}~\cite{opengl}.
To overcome this difficulty, we just set the \opengl{} texture format for the image memory so that it still has 64-bit per pixel and, at the same time, multiple colour channels~(for instance, 16-bit RGBA).
This approach hinders, to some extent, the data access by GPU, but we will show how to deinterleave colour channels to obtain the initial value in Section~\ref{s.storage}.

Note, that the endianness is also accounted by \opengl{}.
FITS stores its data in 8-bit bytes using the big-endian format~\cite{pence_etal2010}.
On the other hand, \opengl{} uses the machine-specific byte ordering~\cite{opengl}.
Today, the vast majority of workstations use the little-endian format, e.g. x86/x86-64~\cite{intel_x86_2015}, ARM machines with Windows~\cite{windows_arm_2016}, iOS~\cite{apple_ios_2013}, and almost all Linux-based operating systems with very few exceptions.
Fortunately, we do not need to explicitly handle the FITS endianness.
It is sufficient to specify that the data are big-endian on CPU when calling the \opengl{} data transport function.
Although it is not strictly specified, the byte swapping may be even performed in the graphics processing unit hardware if this is required while texture storing.

Since the GPU has an access to the FITS data, the image can be easily transformed and rendered.
Note also that the following two tasks are completely separated when programming \opengl{}.
The first task is the geometric transformation of 3D space objects.
Assume that we have a rectangular plane has the same aspect ratio as the initial FITS image has.
A broad variety of geometric transformations may be applied to the plane independently of what is going to be drawn on this plane.
This is how the pan, scale and rotation are implemented.
One only needs to carefully program straightforward transformation formulae given in~\ref{s.coordinate}.

The second task is rendering the texture onto the plane.
This is always performed in the local plane coordinates independently of the final plane orientation and the scale.
The task that is handled in the fragment shader  is essentially the mapping between a local plane coordinate and its visible colour.
Specific coordinates are implicitly determined by \opengl{} and the colours are evaluated simultaneously.

\section{Fragment Shader Organisation}\label{s.storage}
\begin{figure}[t]
\begin{tikzpicture}
\path (2cm, 0.75cm) node [] {\tiny{$(k-1)$-th pixel}}
      ++(4cm, 0) node [] {\tiny{$k$-th pixel}}
      ++(4cm, 0) node [] {\tiny{$(k+1)$-th pixel}};

\draw [decorate,decoration={brace,amplitude=0.125cm}] (0, 0.5cm) -- ++(4cm, 0);
\draw [decorate,decoration={brace,amplitude=0.125cm}] (4cm, 0.5cm) -- ++(4cm, 0);
\draw [decorate,decoration={brace,amplitude=0.125cm}] (8cm, 0.5cm) -- ++(4cm, 0);

\draw[step=.5cm,xshift=0cm] (-0.125cm, 0) grid (12.125cm, 0.5cm);

\path (0.25cm, 0.25cm) node [] {\small{\tt{01}}}
      ++(0.5cm, 0) node [] {\small{\tt{ab}}}
      ++(0.5cm, 0) node [] {\small{\tt{90}}}
      ++(0.5cm, 0) node [] {\small{\tt{7d}}}
      ++(0.5cm, 0) node [] {\small{\tt{60}}}
      ++(0.5cm, 0) node [] {\small{\tt{95}}}
      ++(0.5cm, 0) node [] {\small{\tt{40}}}
      ++(0.5cm, 0) node [] {\small{\tt{00}}}
      ++(0.5cm, 0) node [] {\small{\tt{01}}}
      ++(0.5cm, 0) node [] {\small{\tt{64}}}
      ++(0.5cm, 0) node [] {\small{\tt{92}}}
      ++(0.5cm, 0) node [] {\small{\tt{37}}}
      ++(0.5cm, 0) node [] {\small{\tt{36}}}
      ++(0.5cm, 0) node [] {\small{\tt{7b}}}
      ++(0.5cm, 0) node [] {\small{\tt{68}}}
      ++(0.5cm, 0) node [] {\small{\tt{a0}}}
      ++(0.5cm, 0) node [] {\small{\tt{01}}}
      ++(0.5cm, 0) node [] {\small{\tt{82}}}
      ++(0.5cm, 0) node [] {\small{\tt{fd}}}
      ++(0.5cm, 0) node [] {\small{\tt{f1}}}
      ++(0.5cm, 0) node [] {\small{\tt{55}}}
      ++(0.5cm, 0) node [] {\small{\tt{78}}}
      ++(0.5cm, 0) node [] {\small{\tt{c9}}}
      ++(0.5cm, 0) node [] {\small{\tt{00}}};

\fill[red!30!white] (0, -0.75cm) rectangle +(1cm,0.5cm);
\fill[green!40!white] (1cm, -0.75cm) rectangle +(1cm,0.5cm);
\fill[blue!40!white] (2cm, -0.75cm) rectangle +(1cm,0.5cm);
\fill[black!70!white] (3cm, -0.75cm) rectangle +(1cm,0.5cm);
\fill[red!30!white] (4cm, -0.75cm) rectangle +(1cm,0.5cm);
\fill[green!40!white] (5cm, -0.75cm) rectangle +(1cm,0.5cm);
\fill[blue!40!white] (6cm, -0.75cm) rectangle +(1cm,0.5cm);
\fill[black!70!white] (7cm, -0.75cm) rectangle +(1cm,0.5cm);
\fill[red!30!white] (8cm, -0.75cm) rectangle +(1cm,0.5cm);
\fill[green!40!white] (9cm, -0.75cm) rectangle +(1cm,0.5cm);
\fill[blue!40!white] (10cm, -0.75cm) rectangle +(1cm,0.5cm);
\fill[black!70!white] (11cm, -0.75cm) rectangle +(1cm,0.5cm);

\draw[xshift=0.0cm,yshift=-0.75cm] (-0.125cm, 0cm) grid [step=.5cm] (12.125cm, 0.5cm);

\path (0.25cm, -0.5cm) node [] {\small{\tt{ab}}}
      ++(0.5cm, 0) node [] {\small{\tt{01}}}
      ++(0.5cm, 0) node [] {\small{\tt{7d}}}
      ++(0.5cm, 0) node [] {\small{\tt{90}}}
      ++(0.5cm, 0) node [] {\small{\tt{95}}}
      ++(0.5cm, 0) node [] {\small{\tt{60}}}
      ++(0.5cm, 0) node [white] {\small{\tt{00}}}
      ++(0.5cm, 0) node [white] {\small{\tt{40}}}
      ++(0.5cm, 0) node [] {\small{\tt{64}}}
      ++(0.5cm, 0) node [] {\small{\tt{01}}}
      ++(0.5cm, 0) node [] {\small{\tt{37}}}
      ++(0.5cm, 0) node [] {\small{\tt{92}}}
      ++(0.5cm, 0) node [] {\small{\tt{7b}}}
      ++(0.5cm, 0) node [] {\small{\tt{36}}}
      ++(0.5cm, 0) node [white] {\small{\tt{a0}}}
      ++(0.5cm, 0) node [white] {\small{\tt{68}}}
      ++(0.5cm, 0) node [] {\small{\tt{82}}}
      ++(0.5cm, 0) node [] {\small{\tt{01}}}
      ++(0.5cm, 0) node [] {\small{\tt{f1}}}
      ++(0.5cm, 0) node [] {\small{\tt{fd}}}
      ++(0.5cm, 0) node [] {\small{\tt{78}}}
      ++(0.5cm, 0) node [] {\small{\tt{55}}}
      ++(0.5cm, 0) node [white] {\small{\tt{00}}}
      ++(0.5cm, 0) node [white] {\small{\tt{c9}}};

\draw [decorate,decoration={brace,amplitude=0.125cm,mirror}] (0, -0.75cm) -- ++(1cm, 0);
\draw [decorate,decoration={brace,amplitude=0.125cm,mirror}] (1cm, -0.75cm) -- ++(1cm, 0);
\draw [decorate,decoration={brace,amplitude=0.125cm,mirror}] (2cm, -0.75cm) -- ++(1cm, 0);
\draw [decorate,decoration={brace,amplitude=0.125cm,mirror}] (3cm, -0.75cm) -- ++(1cm, 0);
\draw [decorate,decoration={brace,amplitude=0.125cm,mirror}] (4cm, -0.75cm) -- ++(1cm, 0);
\draw [decorate,decoration={brace,amplitude=0.125cm,mirror}] (5cm, -0.75cm) -- ++(1cm, 0);
\draw [decorate,decoration={brace,amplitude=0.125cm,mirror}] (6cm, -0.75cm) -- ++(1cm, 0);
\draw [decorate,decoration={brace,amplitude=0.125cm,mirror}] (7cm, -0.75cm) -- ++(1cm, 0);
\draw [decorate,decoration={brace,amplitude=0.125cm,mirror}] (8cm, -0.75cm) -- ++(1cm, 0);
\draw [decorate,decoration={brace,amplitude=0.125cm,mirror}] (9cm, -0.75cm) -- ++(1cm, 0);
\draw [decorate,decoration={brace,amplitude=0.125cm,mirror}] (10cm, -0.75cm) -- ++(1cm, 0);
\draw [decorate,decoration={brace,amplitude=0.125cm,mirror}] (11cm, -0.75cm) -- ++(1cm, 0);

\path (0.5cm, -1.0cm) node (red1) [] {\tiny{Red}}
      ++(1cm, 0) node (green1) [] {\tiny{Green}}
      ++(1cm, 0) node (blue1) [] {\tiny{Blue}}
      ++(1cm, 0) node (alpha1) [] {\tiny{Alpha}};

\path (1cm, -1.5cm) node (sred1) [] {\tiny{$6.515 \cdot 10^{-3}$}}
      ++(0cm, -0.25cm) node (sgreen1) [] {\tiny{$5.644 \cdot 10^{-1}$}}
      ++(0cm, -0.25cm) node (sblue1) [] {\tiny{$3.772 \cdot 10^{-1}$}}
      ++(0cm, -0.25cm) node (salpha1) [] {\tiny{$2.500 \cdot 10^{-1}$}};

\draw [{Circle[length=2pt]}-{Circle[length=2pt]}] (red1.south) .. controls (green1.south) .. (sred1.east);
\draw [{Circle[length=2pt]}-{Circle[length=2pt]}] (green1.south) .. controls (blue1.south west) .. (sgreen1.east);
\draw [{Circle[length=2pt]}-{Circle[length=2pt]}] (blue1.south) -- (sblue1.east);
\draw [{Circle[length=2pt]}-{Circle[length=2pt]}] (alpha1.south) -- (salpha1.east);

\path (4.5cm, -0.825cm) node (red2) [] {}
      ++(1cm, 0) node (green2) [] {}
      ++(1cm, 0) node (blue2) [] {}
      ++(1cm, 0) node (alpha2) [] {};

\path (5cm, -1.5cm) node [] (sred2) {\tiny{$5.432 \cdot 10^{-3}$}}
      ++(0cm, -0.25cm) node [] (sgreen2) {\tiny{$5.711 \cdot 10^{-1}$}}
      ++(0cm, -0.25cm) node [] (sblue2) {\tiny{$2.128 \cdot 10^{-1}$}}
      ++(0cm, -0.25cm) node [] (salpha2) {\tiny{$4.087 \cdot 10^{-1}$}};

\draw [{Circle[length=2pt]}-{Circle[length=2pt]}] (red2.south) .. controls (green2.south) .. (sred2.east);
\draw [{Circle[length=2pt]}-{Circle[length=2pt]}] (green2.south) .. controls (blue2.south west) .. (sgreen2.east);
\draw [{Circle[length=2pt]}-{Circle[length=2pt]}] (blue2.south) -- (sblue2.east);
\draw [{Circle[length=2pt]}-{Circle[length=2pt]}] (alpha2.south) -- (salpha2.east);

\path (8.5cm, -0.825cm) node (red3) [] {}
      ++(1cm, 0) node (green3) [] {}
      ++(1cm, 0) node (blue3) [] {}
      ++(1cm, 0) node (alpha3) [] {};

\path (9cm, -1.5cm) node [] (sred3) {\tiny{$5.889 \cdot 10^{-3}$}}
      ++(0cm, -0.25cm) node [] (sgreen3) {\tiny{$9.919 \cdot 10^{-1}$}}
      ++(0cm, -0.25cm) node [] (sblue3) {\tiny{$3.338 \cdot 10^{-1}$}}
      ++(0cm, -0.25cm) node [] (salpha3) {\tiny{$7.851 \cdot 10^{-1}$}};

\draw [{Circle[length=2pt]}-{Circle[length=2pt]}] (red3.south) .. controls (green3.south) .. (sred3.east);
\draw [{Circle[length=2pt]}-{Circle[length=2pt]}] (green3.south) .. controls (blue3.south west) .. (sgreen3.east);
\draw [{Circle[length=2pt]}-{Circle[length=2pt]}] (blue3.south) -- (sblue3.east);
\draw [{Circle[length=2pt]}-{Circle[length=2pt]}] (alpha3.south) -- (salpha3.east);

\path (-0.75cm, 0.25cm) node [align=center] {\tiny{FITS data}}
      ++(0, -0.75cm) node [align=center] {\tiny{Texture}}
      ++(0, -1.375cm) node [align=center] {\tiny{Sampler output}};

\end{tikzpicture}
\caption{\label{fig:memory}Memory layout example.
At the top layer, FITS image linear memory data represent 64-bit integer pixels in big-endian order.
At the middle layer, the texture memory representation for a 16-bit RGBA~(Red, Green, Blue, Alpha) pixel in little-endian architecture is given.
At the bottom layer, floating point vectors are shown that are returned by the sampler when the texture is accessed.
}
\end{figure}
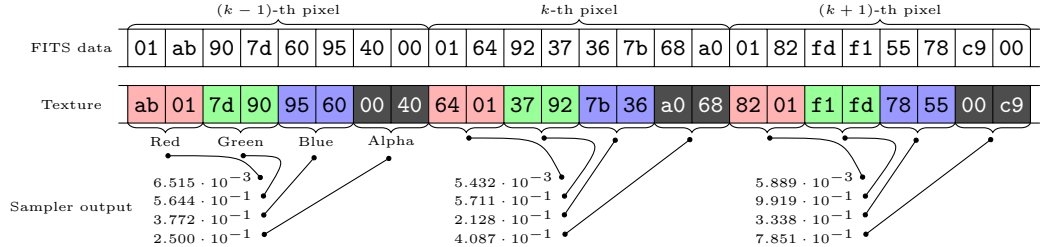
The fragment shader is a place where colour transformations are performed.
Let us describe the tasks to be solved in the shader code:
\begin{itemize}
\item{
FITS format requires the following transformation to be applied to the data stored in the file:
\begin{equation}
\label{eq:fits}
\mathtt{physical\_value} = \mathtt{BSCALE} \cdot \mathtt{array\_value} + \mathtt{BZERO},
\end{equation}
where $\mathtt{BZERO}$ and $\mathtt{BSCALE}$ are stored in the FITS header, $\mathtt{physical\_value}$ is a `true' value, and $\mathtt{array\_value}$ is a byte-representation stored in the file.
Since we load the whole file into the graphics memory, we have to handle this equation using the GPU as well.
}
\item{
Astronomical data usually have a greater number of bits per colour than modern screens have or human eye can distinguish.
In common modern operating systems, 8-bit colours are used, whereas astronomical data coming from CCD are likely to contain 16-bit (or even greater) ones.
This is why we prefer to adjust minimal and maximal levels to give an image more contrast in order to see all interesting details.
}
\item{
Different colour mapping schemes should be applied to data.
In FITS files, we usually have single-channel colour formats.
Each pixel is described by a single value.
 In optical astronomy, it is usually expressed in ADU units (analogue-digital units) coming from a CCD (charge-coupled device).~\cite{Howell2006}
 So it is natural to represent the data as greyscale pixels.
 However, we could use any other colour instead of white to represent the brightest image pixels.
}
\item{
The last but not the least is format converting.
FITS supports different data storage bit depths specified by the $\mathtt{BITPIX}$ header value: up to 64-bit integer and floating point numbers.\footnote{The following FITS data formats exist: 8-bit unsigned integers ($\mathtt{BITPIX} = 8$), 16-, 32-, 64-bit signed integers ($\mathtt{BITPIX}=16,32,64$), single and double precision floating point numbers ($\mathtt{BITPIX}=-32, -64$).}
The challenge here is that earlier \opengl{} versions allow using only 8~or~16~bits per colour channel, and even the latest \opengl{} does not support 64-bit textures.
}
\end{itemize}

To solve this last task, we use the process of deinterleaving colours.
First, we are going to describe how the data are transformed in a forward way in the example shown in~Fig.~\ref{fig:memory}.
After that, we will construct the inverse transformation.
At the top level of Fig.~\ref{fig:memory}, three consecutive pixels of a 64-bit FITS image are shown.
The middle level presents the same data loaded into the little-endian GPU.
Since there is no option to specify its real 64-bit format in \opengl{}, the pixel is considered to have four subsequent 16-bit channels.
The bottom level displays what we may obtain when accessing the texture by the sampler.
These are four floating point numbers to represent a pair of bytes each.
To summarise, each texture pixel for the $i$-th colour channel ($t_i$) may be expressed as follows:
\begin{equation}
	\label{eq:y_to_t}
	t_i = \frac{\word{y}{i}{n}}{2^n - 1}, ~~~ i=0 ... \left(\frac{N}{n}-1\right)\,,
\end{equation}
where $y$ is the initial value;
$N$ is the real bit size of the input FITS pixel~($64$~in~Fig.~\ref{fig:memory});
$n$ is the bit size of the colour channel~($16$~in~Fig.~\ref{fig:memory});
$\word{\cdot}{i}{n}$ denotes the $i$-th $n$-bit word of $\cdot$ value~
(the $i$-th colour channel in Fig.~\ref{fig:memory});
$i = 0$ corresponds to the most significant word.
Since the sampler returns the normalised value, we have $2^n - 1$ in the denominator, while $t_i \in [0;1]$.
Note that Eq.~(\ref{eq:y_to_t}) is still valid when $N=n$ and no colour  transformation is performed.

The expression for the $\mathtt{array\_value}$ $y$ may be written as follows:
\begin{equation}
	\label{eq:t_to_y}
	y = \sum_{i=0}^{\frac{N}{n}-1}{ {2^{N-(i+1)n}} \left(2^{n}-1\right) } \cdot t_i \,.
\end{equation}
Since the internal texture representation is unsigned, the following substitution should be done for the signed integer ($\mathtt{BITPIX}=16, 32, 64$):
\begin{equation}
	\label{eq:t_sing}
	t_0 \leftarrow t_0 - \frac{2^{n}}{2^n-1} \quad \mbox{if} \quad t_0 > \frac{1}{2}\,.
\end{equation}

Using Eq.~(\ref{eq:fits}) and the linear normalisation transformation, we obtain the following expression:
\begin{equation}
	\label{eq:t_to_y_2}
	g = \frac{\mathtt{BSCALE}}{M-m}\sum_{i=0}^{\frac{N}{n}-1}{ {2^{N-(i+1)n}} \left(2^{n}-1\right) } \cdot t_i + \frac{\mathtt{BZERO} - m}{M -m}\,,
\end{equation}
where $m$ and $M$ are the minimal and maximal physical values set by user, respectively, according to Eq.~(\ref{eq:fits}). If a $\mathtt{physical\_value} \in [m;M]$, then $g \in [0;1]$, otherwise $g$ is clamped to this interval.

Equation~(\ref{eq:t_to_y_2}) may be rewritten in the following convenient inner product form:
\begin{equation}
	\label{eq:t_to_y_vec}
	g = \left(\mathbf{c}, \mathbf{t} - \mathbf{z}\right),
\end{equation}
where $\mathbf{t}$ is the vector of $t_i$ values, $\mathbf{c}$ and $\mathbf{z}$ are vector functions of $\mathtt{BSCALE}$, $\mathtt{BZERO}$, $m$, and $M$.
The vectors $\mathbf{c}$ and $\mathbf{z}$ are constant for every image pixel, so they can be precalculated on the CPU. \opengl{} allows us to efficiently calculate inner products on the GPU.
Note that by construction, $\left|z_i\right|<1$ for every~$i$, what ensures that the precision is not lost in numerical operations.

The final fragment colour is the $\mathbf{f}$ vector consisting of four normalised components: red, green, blue, and alpha.
To represent the data in user colours, we may calculate the fragment colour as follows:
\begin{equation}
	\mathbf{f} = \left(1-g\right) \mathbf{f_{0}} + g \mathbf{f_{1}},
\end{equation}
since $g$ is implicitly clamped to the $[0;1]$ interval after using Eq.~(\ref{eq:t_to_y_vec}).
This linear interpolation is implicitly performed by the sampler of special one-dimensional texture consisting of two-colour pixels (the interpolation node) that store the $\mathbf{f_0}$ and $\mathbf{f_1}$ colour constants.
Further generalisation may be  achieved if additional interpolation nodes are added to the colour map texture, allowing an easy implementation of multi-colour maps such as `black--blue--yellow`.

\begin{table}[t]
\begin{center}
\begin{tabular}{c|l|l}
FITS $\mathtt{BITPIX}$ & \opengl{}~2.1 & \opengl{}~3 \\
\hline
$8$ & $n=8$ (native) & $n=8$ (native)\\
$16$ & $n=8$ (Luminance, Alpha) & $n=16$ (native)\\
$32$ & $n=8$ (RGBA)& $n=16$ (Red, Green) \\
$64$ & $n=16$ (RGBA)& $n=16$ (RGBA) \\
$-32$ & not supported$^{\dagger}$ & native \\
$-64$ & not supported & native via the extension$^{*}$ \\
\end{tabular}
\end{center}
\caption{\label{table:bitpix}Support matrix.
Here, the specific values of $n$ from Eq.~(\ref{eq:y_to_t}) are presented for different possible cases.
$N=\mathtt{BITPIX}$ for integer formats, negative $\mathtt{BITPIX}$ denotes floating point numbers. The case $n=N$, when  no colour trick is used, is denoted as `native'.
\newline
$^{\dagger}$single precision floating point textures are supported in \opengl{}~2.1 via extensions but cannot be used in \fips{} due to Qt framework limitations.
\newline
$^{*}$double precision floating point values must be unpacked from two 32-bit unsigned integers using {\tt ARB\_gpu\_shader\_fp64} extension.
}
\end{table}

In Table~\ref{table:bitpix} we present particular values for $n$ in different possible cases.
Note that the following equation should be used instead of~(\ref{eq:t_to_y_2}) for floating point formats:
\begin{equation}
	\label{eq:t_to_y_f}
	g = \frac{\mathtt{BSCALE}}{M-m} y + \frac{\mathtt{BZERO} - m}{M -m}\,,
\end{equation}
because the sampler output $y$ is not normalised for floating point textures, and the colour trick is not used in this specific case of floating point number representation in memory.
In order to handle double precision floating point FITS files, {\tt ARB\_gpu\_shader\_fp64} extension should be used~\cite{arb_gpu_shader_fp64}.
In this case the internal representation of GPU data is set up to two-channel 32-bit unsigned integers.
After that, the sampler output in the fragment shader is subjected to the action of a specific built-in magic function.
The function performs reinterpretation cast, taking a pair of 32-bit unsigned integers obtained from the texture, and returns the double precision floating point value.

\section{Software Evaluation and Validation}\label{s.eval}
\begin{figure}[t!]
 \centering
\begin{tabular}{cc}
 \includegraphics[width=6cm]{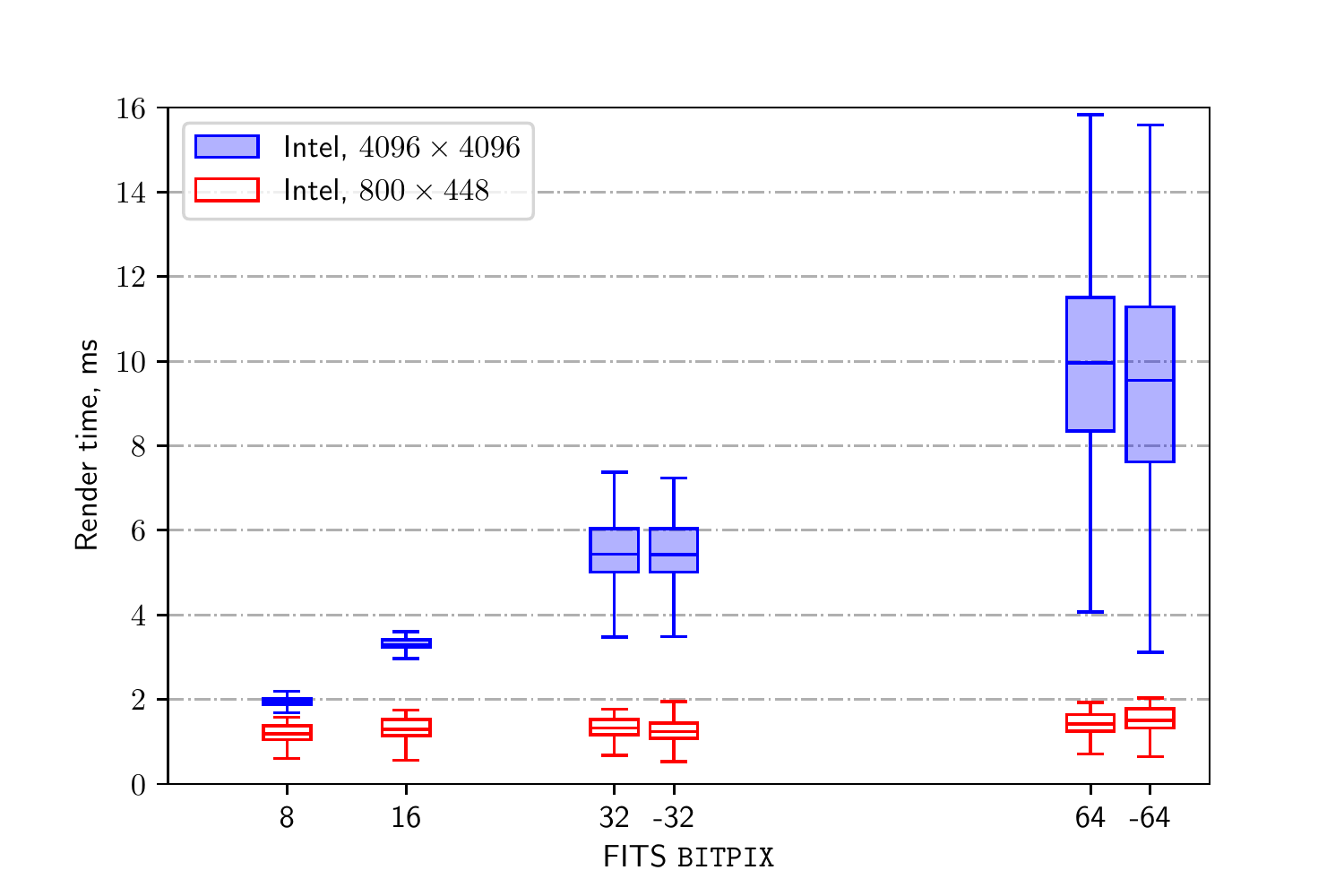} &
 \includegraphics[width=6cm]{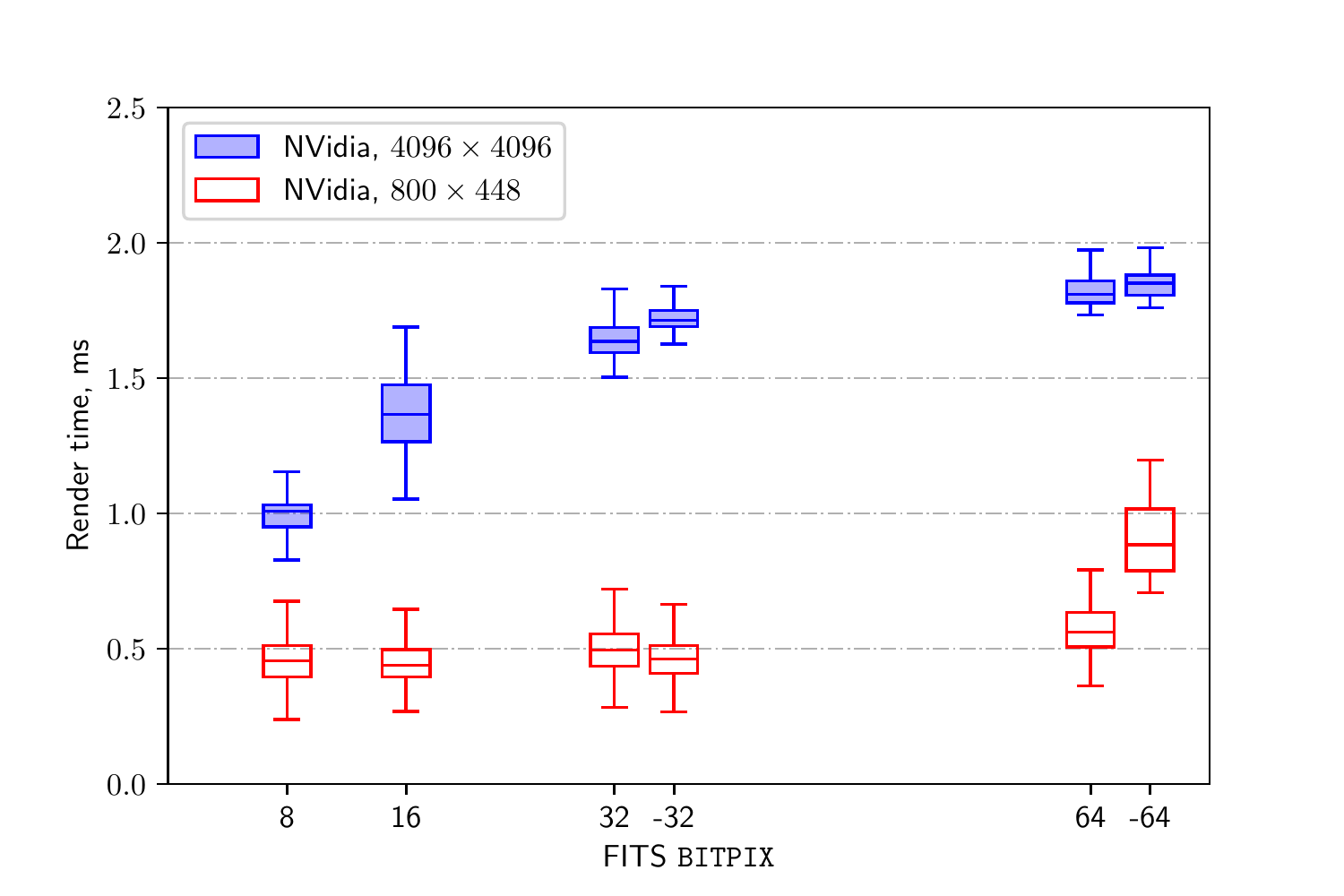}
\end{tabular}
 \caption{\label{fig:eval_hw}
Box-and-whisker plots for rendering time at Intel HD~Graphics~4000~(left panel) and NVidia GeForce~GT~635M~(right panel).
The boxes show the quartiles.
The ends of the whiskers represent the lowest and the highest result still within $1.5$ interquartile range.
}
\end{figure}

\begin{figure}[t!]
 \centering
 \includegraphics[width=6cm]{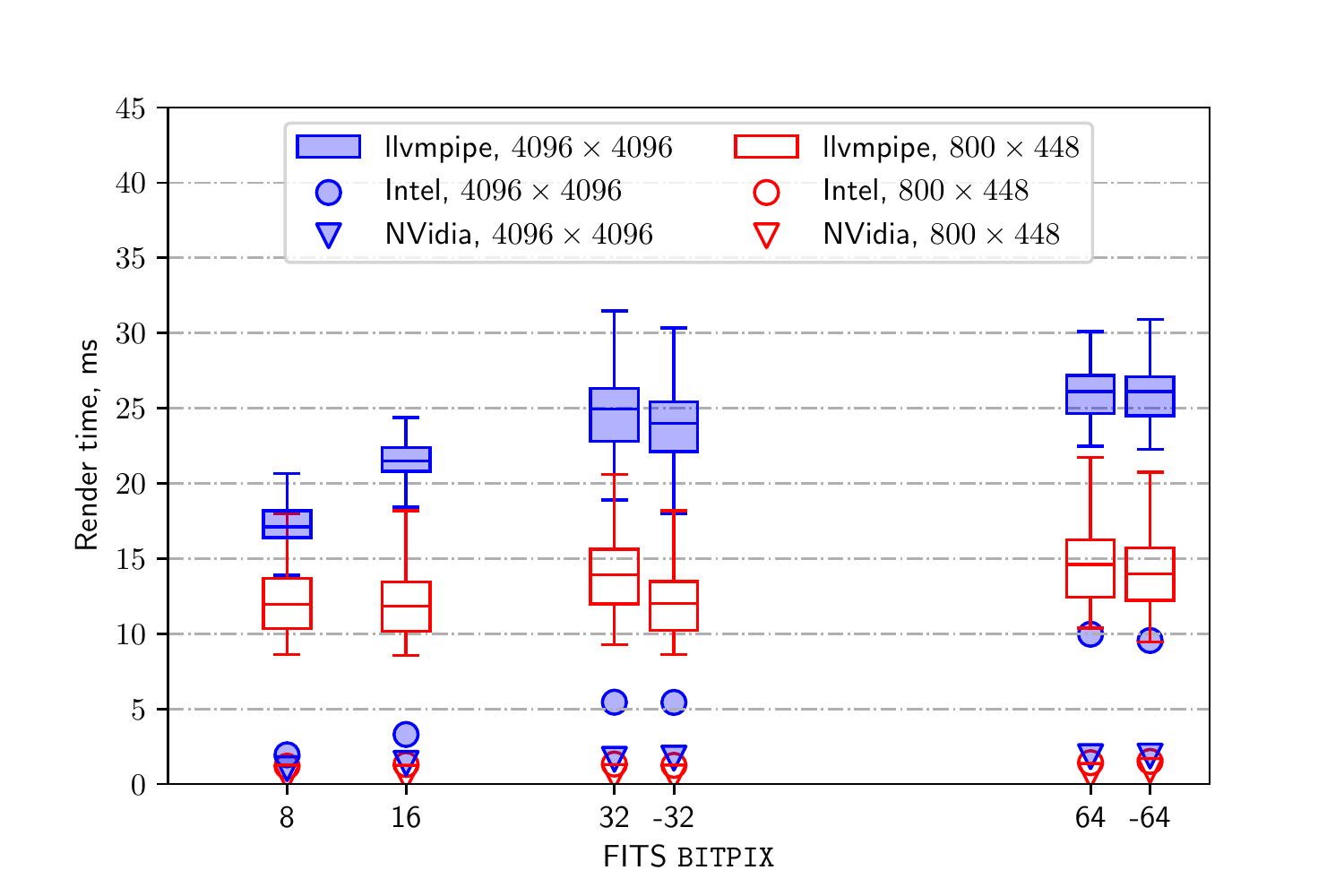}
 \caption{\label{fig:eval}
Box-and-whisker plot for rendering time at Mesa llvmpipe software renderer.
The notation is the same as for Fig~\ref{fig:eval_hw}.
Median rendering times for Intel and NVidia are shown for comparison.
}
\end{figure}


Users expect modern applications to provide responsive user interface.
Interface responsiveness gives a user ability to immediately see the results of user actions.
For instance, it is important when touch-pad gestures are employed as a part of user interface requiring continuous user actions to be handled with low possible latency.
In our case the following user actions trigger the image redrawing: changing the cut levels, switching colour maps, zooming, panning, and rotating the image.
We measured the time required to redraw the image under these typical actions.

Three hardware setups have been tested with the laptop PC.
The software rendering was performed using Mesa llvmpipe\footnote{\url{https://www.mesa3d.org/llvmpipe.html}} driver with Intel Core~i7-3520M CPU, the hardware rendering was performed  using integrated Intel HD~Graphics~4000, and the hardware rendering was performed using discrete NVidia GeForce~GT~635M.
We considered two image geometry sizes: $800 \times 448$ pixels and $4096 \times 4096$ pixels, with all six bit depths,
twelve image files in total.
In all cases the application ran in full screen mode at $1980 \times 1080$ resolution.
This allowed us to examine two cases: when the image was stretched for the small images and shrunk for the large images.
For each case, a few hundreds of drawings were carried out.

We found that the image rotation rendering is the most expensive of all operations, so expenses for it may be considered as an upper bound estimate compared to other expenses.
The obtained results are given in Fig.~\ref{fig:eval_hw} and Fig.~\ref{fig:eval}.
The following specific features can be seen in the figures.
One may see that Intel hardware behaves similarly to NVidia hardware.
Note that the required memory size is proportional to the image bit depth.
In Fig.~\ref{fig:eval_hw}, one can see that the rendering time for the $4096 \times 4096$ pixels image depends almost linearly on the bit depth.
 In this case, the performance seems to be limited mainly by the memory bandwidth.
At the same time, the $800 \times 448$ pixels image is small enough to fit into modern memory caches and therefore no dependence on the bit depth is revealed in all these three rendering engines.
Also in the case when the double precision floating point image is formed on the NVidia GPU, some specific additional overheads appear.

The same kind of measurements were carried out for panning.
We found that the rendering time doesn't depend on the image bit depth.
It is approximately the same as the time for rotating the 8-bit image.
It seems that the time depends only on the visible image size.

The llvmpipe renderer is considerably slower as compared to the hardware cases.
Note that the llvmpipe produces just-in-time compiled code highly optimised for using CPU vector instructions.
Hence, the llvmpipe rendering time is considered as a lower bound estimate for any CPU-based implementation.
However, we were able to see some glitches during llvmpipe rendering for all considered actions.
The glitches look like vertical synchronisation issues: the top and bottom parts of the picture correspond to the different subsequent frames.
This is in agreement with Fig.~\ref{fig:eval}.
Indeed, it may take more than $30$~ms to render the $4096 \times 4096$ pixels image which corresponds to $\approx30$ frames per second, given that the measurements were carried out on an unloaded CPU.
The measuring data strongly indicate that the GPU-based implementation is more efficient.

Functional testing of the graphic user interface application is a more challenging task than to test the server application.
This is explained by the fact that automatic unit tests cannot be applied to the application graphic interface parts.
Therefore, we use the unit testing where it is possible, such as parsing the FITS files, or assessing the coefficients in Eq.~(\ref{eq:t_to_y_vec}).
The integration of Travis CI and GitHub allows us to easily check the application for a wide range of compilers and Qt framework versions.

To ensure validity of the algorithm described in Section~\ref{s.storage},
we may either render the image to a texture framebuffer, extracting pixel numeric values, 
or we may use  the system colour picker application to see pixel values on the screen.
We used the latter approach since our key interest is to find out what the user can actually see on the screen.
A simple greyscale gradient testing image allows us to perform this kind of end-to-end manual testing.
 
\section{Discussion}\label{s.discussion}
The \opengl{} application may be helpful for astronomical purposes not only due to its ability to perform simple FITS~image rendering.
The so-called texture arrays available with \opengl{}~3 may be properly used to support 3D FITS data.
A 3D data cube stored in GPU video memory in the form of texture array representation may give us an opportunity to implement efficient GPU video playback.
The representation of data cubes in the form of a video is currently implemented in e.g. Ginga and FITSWebQL FITS viewers~\cite{jeschke_etal2013,Zapart2018}.

Since  \opengl{} is capable to support the opacity, we may use another interesting option; namely, to draw multiple FITS~images as semi-transparent colour layers, differently placed and differently orientated.
This option would be helpful for applications similar to those in which the blinking technique in astronomy  is currently used: the manual transient object search.
A similar technique is widely adopted by amateur astronomers~\cite{Heafner, Falla}.
However, in this case, a nonlinear geometric transformation of the image is ordinarily required to compensate for third order aberrations.
This may be easily done using the expansion of a current four-vertex-rectangle plane to a triangle mesh, a common technique employed in the 3D computing world.

Note also that embedding such a geometric transformation engine into the astronomical image processing pipeline considerably increases  the processing rate.
This property is very valuable when looking for asteroids using robotic observatories.
The technology incorporating \opengl{} into the head-less server application is called EGL (Embedded GL); it uses a texture attached to a frame buffer object instead of the hardware display that performs scene rendering~\cite{egl}.
Thus, the rendering output may be downloaded to the CPU after the render operation is performed.

One challenge for upcoming extra-large survey projects, such as LSST~\cite{Becker2007}, is to increase the astronomical image processing cadence.
When using \opengl{} computing shaders (available with \opengl{}~4.3) or OpenCL/\opengl{} integration, a full stack of the common astronomical pipeline may be implemented: from bias subtraction and flat fielding to pixel clustering for star extraction, as it has been proposed e.g. by Warner, et al.~\cite{Warner2013}.

Many of the techniques we have described could be applied to WebGL-based FITS image rendering~\cite{webgl}.
WebGL is a standard programming interface for modern web browsers that allows \opengl{} programming in JavaScript.
Several years ago, there was some publication of plans for the implementation of telescope control system interfaces and astronomical data archives in the form of web applications.
This approach has many advantages.
For example, web applications scale well for any OS and any platform, they do not require installation and may be easily run on a guest-observer laptop~\cite{mandel,Roby2016,Zapart2018}.
A possibility to render an initial data image acquired from the hardware is also of importance here.
These are most important cases when the implementation of  WebGL-based application techniques may prove very helpful.

\section{Conclusion}\label{s.conclusion}
In this paper we have introduced new software for rendering astronomical data in the form of FITS images.
The major design novelty is using GPU acceleration: the image geometry and colour transformation are programmed in GPU using the \opengl{} programming interface.
It turns out that the full processing stack, starting with loading bytes from a FITS file into the GPU memory, to rendering the picture on the user screen, may be implemented by applying all necessary data transformations in the GPU.

\opengl{} provides basically two main features: it increases processing speed using hardware specially designed for geometric and colour image
transformations, and simplifies programming such transformations in case the developer needs them.

The proposed design may clearly have  some important effects for astronomy.
For example, a decrease in the required CPU load would obviously improve the experience of the end user.
Some other opportunities that may arise when using \opengl{} to process astronomical data are yet to be investigated.
\fips{} source codes are available at the GitHub web site, so we hope that other open source developers will join the efforts for further software improvement.
Prebuilt binary packages are available for Windows, openSUSE 15.1+, Fedora 30+, Homebrew Caskroom macOS package manager, and Flatpak Linux package manager.

\section*{Acknowledgements}
Authors thank the referees for the constructive comments which helped to improve the paper.
We also thank Ivan Migalev from Polzunov Altai State Technical University (Barnaul, Russia) for the preparation of \fips{} Windows package, and our colleague Maria Pruzhinskaya for useful discussions on the paper.
The study was partially supported by RBFR grants 18-32-00426 (when preparing the paper) and 18-32-00553 (when verifying the fragment shader equations).

\section*{References}

\bibliography{mybibfile}

\appendix
\section{Coordinate Systems}\label{s.coordinate}
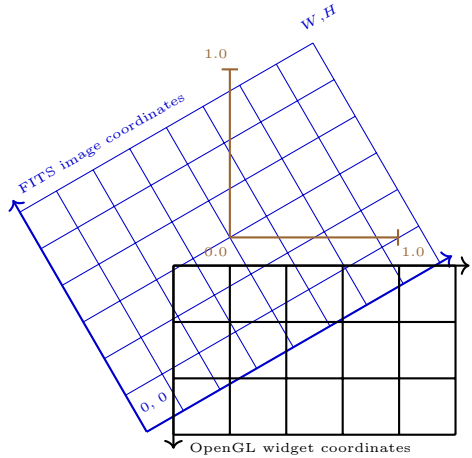
\begin{figure}[t!]
\centering
\begin{tikzpicture}[scale=0.75]
\draw[xshift=-3cm, yshift=-2.25cm, rotate around={30:(3cm,2.25cm)}, color=blue!80!black, step=0.75cm] (0, 0) grid (6cm,4.5cm);
\draw[xshift=-3cm, yshift=-2.25cm, rotate around={30:(3cm,2.25cm)}, color=blue!80!black,thick, ->] (0,0cm) -- (6.25cm,0cm);
\draw[xshift=-3cm, yshift=-2.25cm, rotate around={30:(3cm,2.25cm)}, color=blue!80!black,thick, ->] (0,0cm) -- (0,4.75cm);

\path[xshift=-3cm, yshift=-2.25cm, rotate around={30:(3cm,2.25cm)}]
	(1.875cm, 4.825cm) node [rotate=30,color=blue!80!black] {\tiny{FITS image coordinates}}
	(0.375cm, 0.375cm) node [rotate=30,color=blue!80!black] {\tiny{$0$, $0$}}
      (6.325cm, 4.825cm) node [rotate=30,color=blue!80!black] {\tiny{$W$,$H$}};

\draw[color=brown!80!black,thick,-|] (0,0cm) -- (3cm,0cm);
\draw[color=brown!80!black,thick,-|] (0,0cm) -- (0,3cm);

\path (-0.25cm, -0.25cm) node [color=brown!80!black] {\tiny{0.0}}
      (3.25cm, -0.25cm) node [color=brown!80!black] {\tiny{1.0}}
      (-0.25cm, 3.25cm) node [color=brown!80!black] {\tiny{1.0}};

\path [xshift=-1.0cm, yshift=-3.5cm]
(2.25cm, -0.25cm) node [] {\tiny{\opengl{} widget coordinates}};

\draw[xshift=-1.0cm, yshift=-3.5cm,thick, step=1cm] (0, 0) grid (5cm,3cm);
\draw[xshift=-1.0cm, yshift=-3.5cm,thick, ->] (0,3cm) -- (5.25cm,3cm);
\draw[xshift=-1.0cm, yshift=-3.5cm,thick, ->] (0,3cm) -- (0,-0.25cm);
\end{tikzpicture}
\caption{\label{fig:coordinates}A sample of \fips{} basic coordinate systems.
The tilted rectangle represents a $8\times6$ FITS image, where (0; 0) are the FITS file pixel coordinates of the left bottom.
The erect rectangle represents a $5\times3$ \opengl{} widget, with the origin of coordinates at the top left corner of this rectangle.
The third coordinate system, at the centre of the image, is the world coordinate system, with the origin located at the image centre; the image size equals $(2,3/2)$ in this system of coordinates.
The FITS image is rotated by an angle of $\alpha=30^\circ$~\eqref{eq:rotation}, the widget view rectangle centre has world coordinates $x_{4\mathrm{r}} = 0.5$, $y_{4\mathrm{r}} = 2/3$; the side sizes are $W_\mathrm{r} = 5/3$, $H_\mathrm{r} = 1$~\eqref{eq:widget_view}.}
\end{figure}

At first glance, we deal with at least two different coordinate systems: the FITS image pixel coordinate system and the pixel coordinate system displayed on the user screen.
Since \fips{} can pan, rotate and zoom the image, FITS pixels are transformed to screen pixels in a complicated way, which is described step by step below.
Note that some of these steps are performed implicitly using \opengl{}.
A user can also obtain the coordinates and the value of FITS pixels, so \fips{} can perform all these transformations in the backward direction, from screen coordinates to FITS coordinates.
A sample of \fips{} basic coordinate systems is provided in Fig.~\ref{fig:coordinates}.

\subsection{FITS file pixel coordinates}
It follows from the FITS specification that pixels of a two-dimensional image are numbered by integers, starting from the left bottom corner of the picture~\cite{pence_etal2010}.
Let $x_1$ and $y_1$ denote these coordinates, then $0 \le x_1 < W$ and $0 \le y_1 < H$, where $W$ and $H$ are the image width and height respectively.

\subsection{\opengl{} texture normalised coordinates}
When the image is unpacked to a \opengl{} texture, the internal \opengl{} normalised texture coordinates $x_2$, $y_2$ (the so-called UV-coordinates) are associated with this object.

The coordinate transformation is implicitly performed using \opengl{}.
The inverse transformation is performed as follows:
\begin{equation}
x_1 = \left\lfloor x_2 W \right\rfloor,\quad
y_1 = \left\lfloor y_2 H \right\rfloor,
\end{equation}
where $\left\lfloor \cdot \right\rfloor$ denotes the floor operation, natural for modern processing units when floating point data are converted to integers.

\subsection{Scaled plane coordinates}
To render the texture, one needs a surface.
In our case, the surface consists of four vertices.
In order to keep the initial aspect ratio of the picture, we set the corner coordinates to be $(-Wf,
-Hf)$, $(-Wf,Hf)$, and so on.
Here, $f \equiv \left(\max\left\{H, W\right\}\right)^{-1}$ is a scale factor to retain the plane width and height within a value of $2$.
The inverse transform is given by the following equations:
\begin{equation}
x_2 = \frac{x_3}{2 W f} + \frac{1}{2},\quad
y_2 = \frac{y_3}{2 H f} + \frac{1}{2}.
\end{equation}

\subsection{World coordinates}
The image plane itself is arbitrarily oriented with respect to the zero point of \opengl{} world coordinates.
The orientation is specified by the rotation angle $\alpha$.
\begin{equation}\label{eq:rotation}
x_4 =  x_3 \cos \alpha + y_3 \sin \alpha,\quad
y_4 = -x_3 \sin \alpha + y_3 \cos \alpha.
\end{equation}

\subsection{Widget view coordinates}
Qt \opengl{} widget\footnote{\opengl{} widget is a rectangle part of the application window that shows \opengl{} content} cuts a rectangle from the world coordinate plane.
The ratio of the rectangle sides $W_\mathrm{r}, H_\mathrm{r}$ (in world coordinates) equals the ratio of the widget sides $w, h$.
The origin of the widget view coordinates is in the rectangle centre, while rectangle's bottom right corner has coordinates $(1,1)$ and its top left corner has coordinates $(-1,-1)$.
\begin{equation}\label{eq:widget_view}
x_4 = x_{4\mathrm{r}} + x_5\frac{W_\mathrm{r}}{2},\quad
y_4 = y_{4\mathrm{r}} - y_5\frac{H_\mathrm{r}}{2},
\end{equation}
where $(x_{4\mathrm{r}}, y_{4\mathrm{r}})$ are coordinates of the widget view rectangle in world coordinates.

A smaller widget view rectangle with the size $W_\mathrm{r}, H_\mathrm{r}$ corresponds to a larger zoom-factor.

\subsection{Widget pixel coordinates}
The whole visible \opengl{} world is rendered into the Qt widget, where each pixel is again numbered by an integer, starting from the left top corner.
The transformation is an implicit part of the graphics rendering process performed by the window manager and operating system.
However, the most interesting for us is the explicit form of the inverse transform:
\begin{equation}
x_5 = \frac{2x_6 - \left(w-1\right)}{w},\quad
y_5 = \frac{2y_6 - \left(h-1\right)}{h},
\end{equation}
where, as mentioned above, $w$ and $h$ are the pixel sizes of the \opengl{} widget.
\end{document}